%% file: draft_pII_v06.tex
\documentclass[useAMS,usenatbib]{mn2e}
\input psfig.tex

\title[Galaxy populations in RzCS 052]{Galaxy evolution in
the high redshift, colour--selected
cluster RzCS 052 at $\mathbf{z=1.02}$}
\author[Andreon et al.]{S.
Andreon,$^1$\thanks{stefano.andreon@brera.inaf.it},
E. Puddu$^2$, R. De Propris$^3$, J.-C. Cuillandre$^{4,5}$ \\
$^1$INAF--Osservatorio Astronomico di Brera, via Brera 28, 20121, Milano, 
Italy\\
$^2$INAF--Osservatorio Astronomico di Capodimonte, salita
Moiariello 16, 80131 Napoli, Italy\\
$^3$Cerro Tololo Inter--American Observatory, La Serena, Chile\\
$^4$Canada--France--Hawaii Telescope, PO Box 1597, Kamuela, HI, 96743, USA\\
$^5$Observatoire de Paris, avenue de L'Observatoire 61, 75014 Paris, France\\
}
\date{Accepted ... Received ...}
\pagerange{\pageref{firstpage}--\pageref{lastpage}}
\pubyear{2007}
\begin{document}
\maketitle

\label{firstpage}

\begin{abstract}
We present deep $I$ and $z'$ imaging of the colour-selected
cluster RzCS 052 and study the color-magnitude relation of this cluster,
its scatter, the morphological distribution on the red sequence, the
luminosity and stellar mass functions of red galaxies and the cluster 
blue fraction. We find that the stellar populations of early type galaxies
in this cluster are uniformly old and that their luminosity function does
not show any sign of evolution other than the passive evolution of their
stellar populations. We rule out a significant contribution from 
mergers in the buildup of the red sequence of RzCS 052. The
cluster has a large ($\sim 30\% $) blue fraction and 
and we infer that the evolution of the blue galaxies is faster
than an exponentially declining star formation model and that these objects
have probably experienced starburst episodes.
Mergers are unlikely to be the driver of the observed colour
evolution, because of the measured constancy of
the mass function, as derived from near-infrared photometry of 32
clusters, including RzCS 052, presented in a related paper. 
Mechanisms with clustercentric radial dependent
efficiencies are disfavored as well, because of the
observed constant blue fraction with clustercentric distance.
\end{abstract}

\section{Introduction}

The existence of a tight, and apparently universal, color-magnitude 
relation for galaxies in nearby clusters (e.g. Bower, Lucey \& Ellis
1992; Andreon 2003; 
Lopez-Cruz, Barkhouse \&
Yee 2004; McIntosh, Rix \& Caldwell 2005; Eisenhardt et al. 2007 and
references therein)
implies that the majority of the stellar populations of early-type
cluster galaxies were formed at  $z \gg 1$ over 
relatively short timescales. Studies of 
clusters at high redshift, then, should allow us to witness the 
earlier stages of galaxy evolution, leading to the establishment
of the present
day luminosity function, color-magnitude relation and morphological mixtures. 
A classical example  of this kind of studies 
is the detection of
a blueing trend among galaxies in clusters at $z > 0.3$ by Butcher
\& Oemler (1984).

Until large samples of high redshift ($z \ga 0.8$) clusters become available,
detailed `case studies' of individual objects may provide useful clues
to the evolution of galaxy populations at half the Hubble time and
beyond. Several studies have analyzed
a number of such objects in detail (Blakeslee et al. 2003, 2006; Homeier
et al. 2005, 2006; Holden et al. 2006; Mei et al. 2006a,b), using both 
ground-based imaging and spectroscopy and high-resolution imaging with 
the Advanced Camera for Surveys (ACS) on the Hubble Space Telescope (HST). 
These observations
reiterate that the cluster early-type populations appear to be composed
of old stellar populations which were probably in place at high redshift.

Most of the studied high redshift clusters are selected from the 
X-ray catalogs, which may pre-select objects that have
already formed a deep potential well. An alternative strategy
is to use clusters selected via the prominent red sequence of early
type galaxies (Gladders \& Yee 2000). Several $z\geq 1$ clusters have
already been identified using the galaxy colours or their spectral
energy distributions (Andreon et al. 2005; Stanford et al. 2005).

Here we focus on RzCS 052 (J022143-0321.7), 
a rich (Abell class 2 or 3) cluster selected via a modified red sequence 
method, with a measured redshift of $1.016$ and a velocity
dispersion of $710\pm150$ km s$^{-1}$ and a modest X-ray luminosity of
$(0.68 \pm 0.47) \times 10^{44}$
ergs s$^{-1}$ in the [1-4] keV band. 
Details about this objects and its properties may be
found in Andreon et al. (2007). 
Because RzCS 052 is less X-ray luminous than clusters at similar
redshift, and therefore does not possess a massive X-ray atmosphere, this
object allows us to carry out a study of galaxy evolution in a different
cluster environment and isolate the effects of gas on galaxy properties
(Moran et al. 2007).

The layout of the paper is as follows. We present the data reduction and analysis
in the next section. Section 3 discusses the red sequence galaxies. Section 4 deals
with the blue galaxies in RzCS 052. Finally, we summarize our results in Section 5. 
We adopt the concordance cosmological parameters 
$\Omega_M=0.3$, $\Omega_{\Lambda}=0.7$ and $H_0=70$ km s$^{-1}$ Mpc$^{-1}$. 
Magnitudes 
are quoted in their native photometric system (Vega for $RI$, SDSS for $z'$ and instrumental 
for Megacam data).
Results of our stochastical computations are quoted in
the form $x\pm\sigma$ where $x$ is the posterior
mean and $\sigma$ is the posterior standard deviation.

\begin{figure}
\hbox{%
\psfig{figure=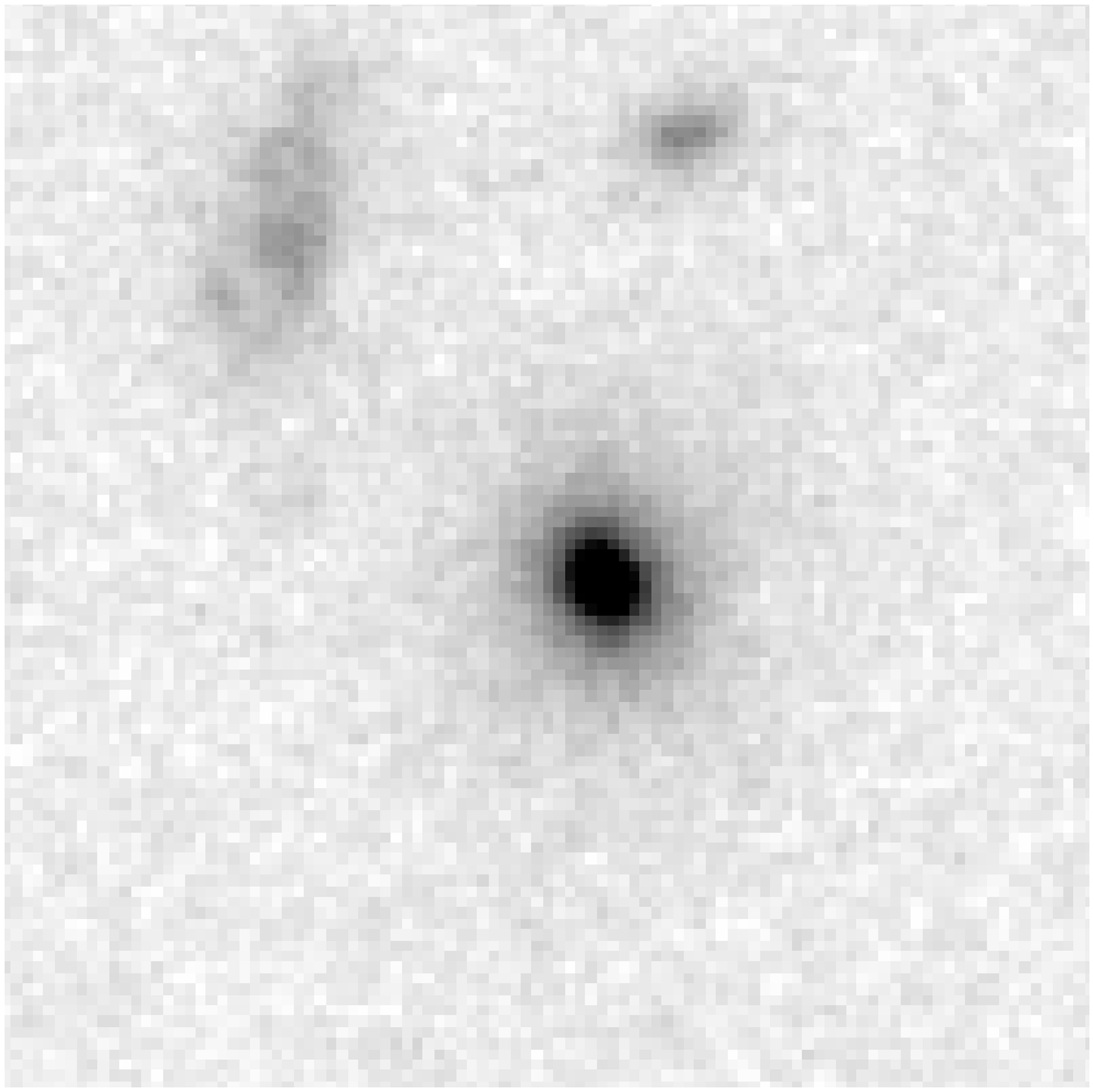,width=2truecm,clip=}%
\psfig{figure=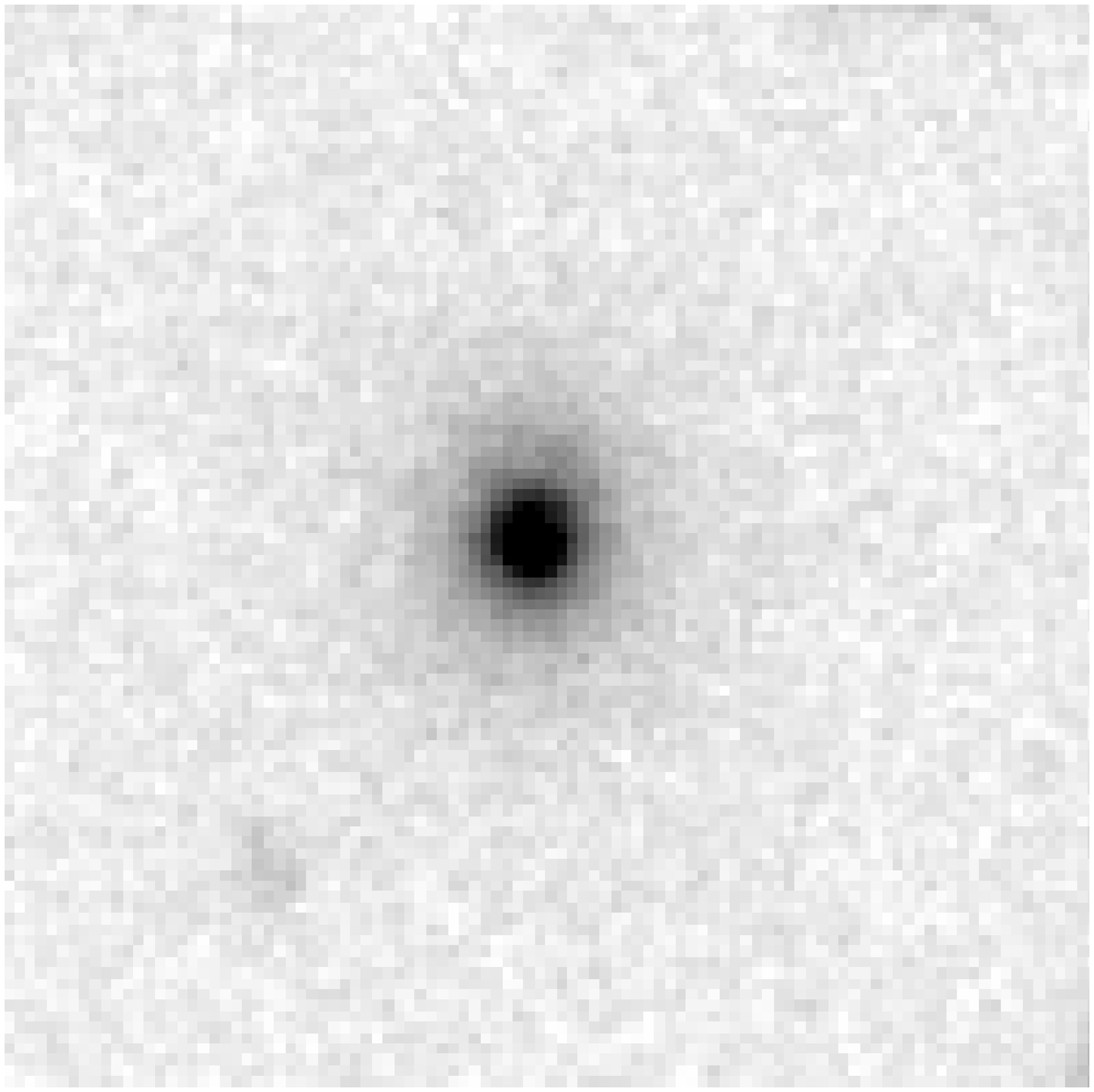,width=2truecm,clip=}%
\psfig{figure=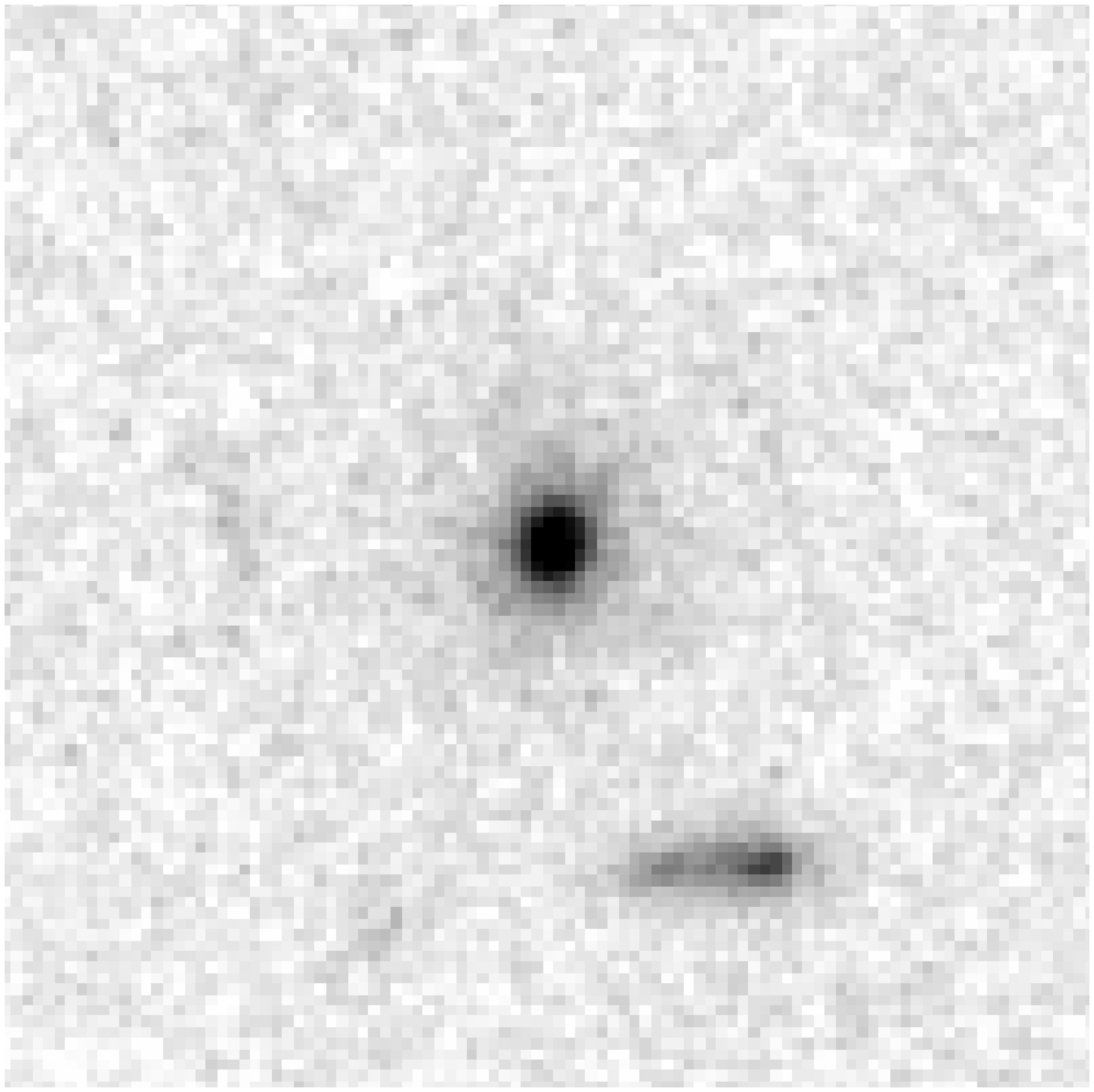,width=2truecm,clip=}%
}
\hbox{%
\psfig{figure=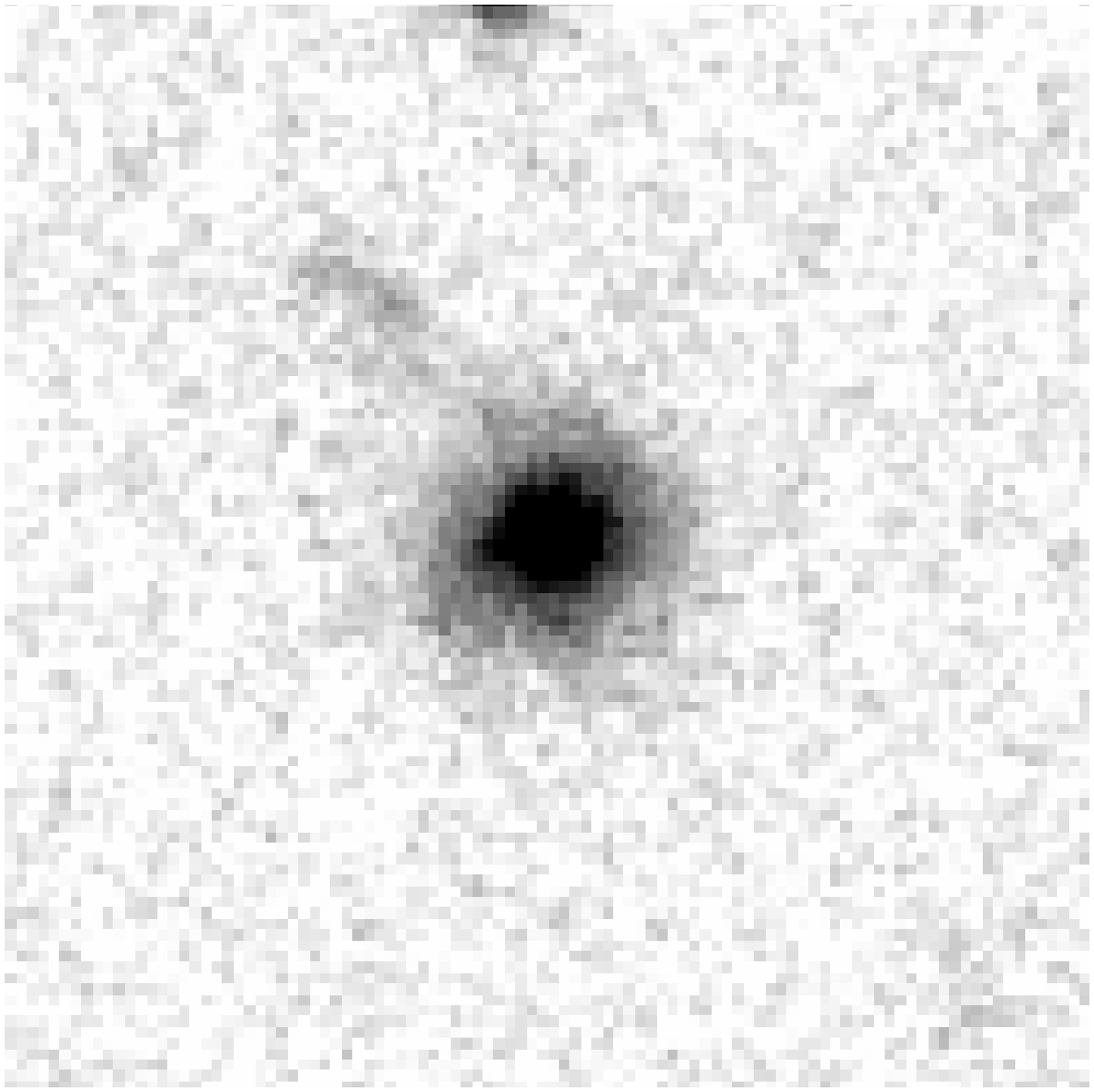,width=2truecm,clip=}%
\psfig{figure=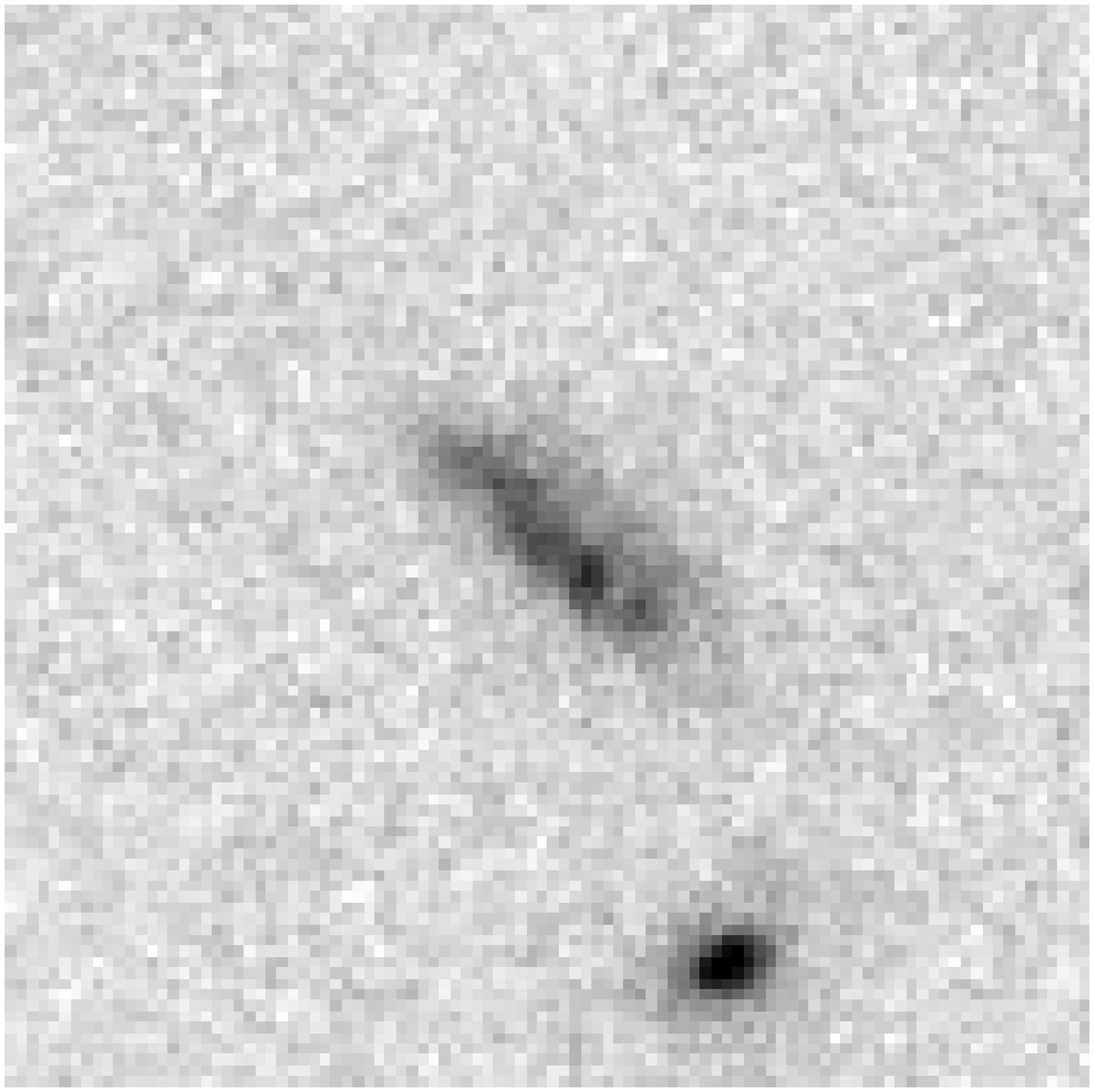,width=2truecm,clip=}%
\psfig{figure=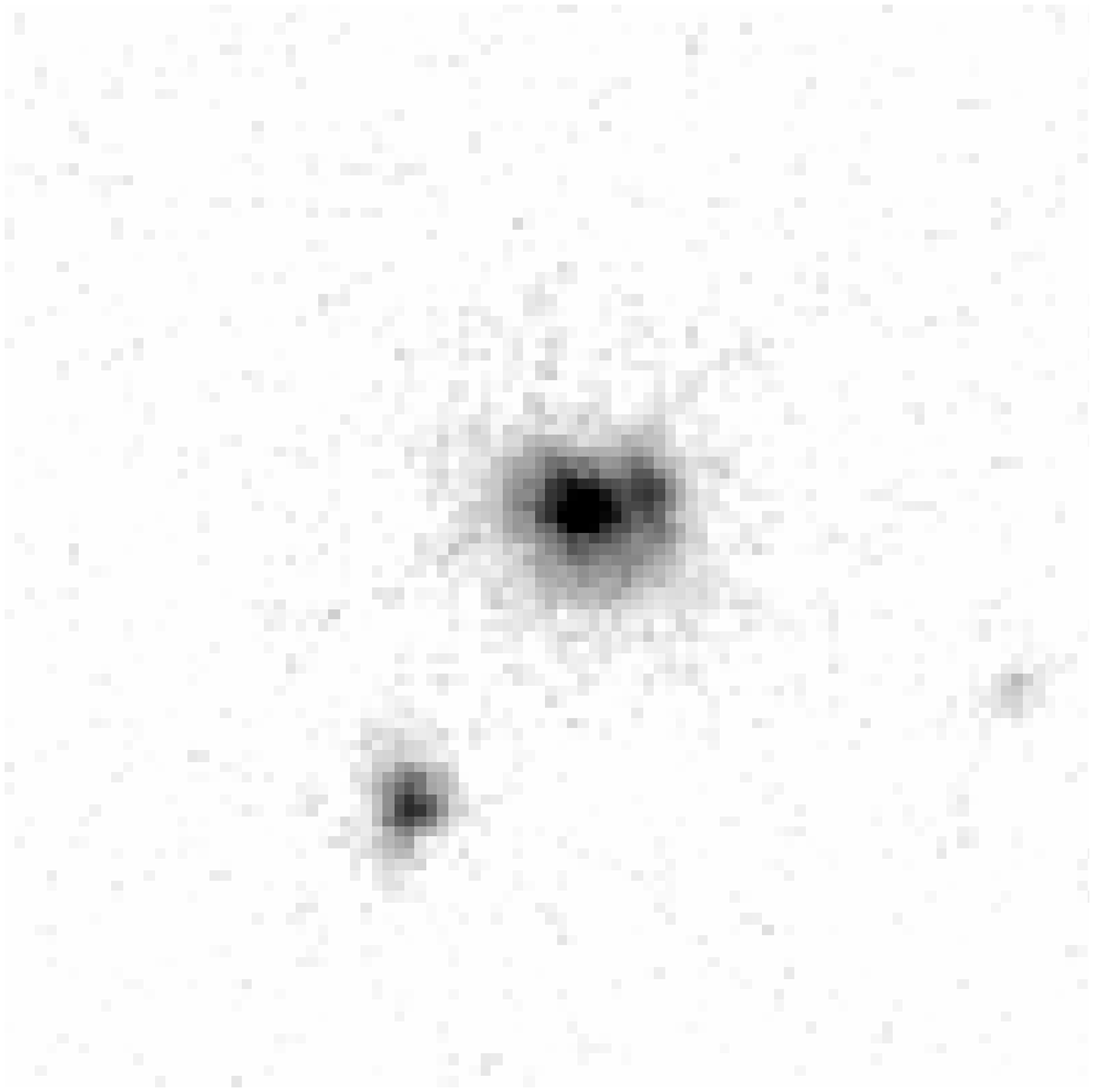,width=2truecm,clip=}%
\psfig{figure=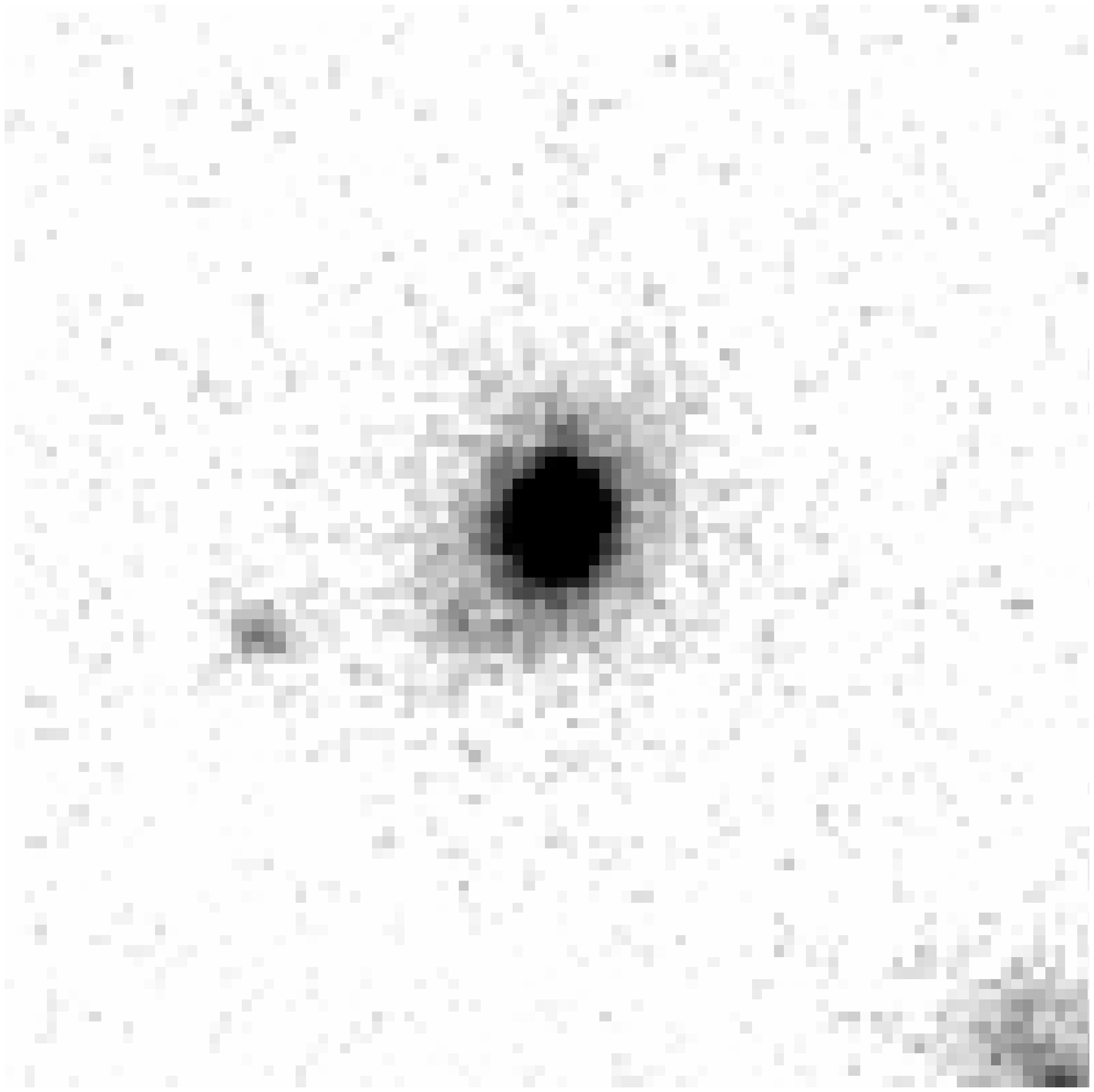,width=2truecm,clip=}%
}
\hbox{%
\psfig{figure=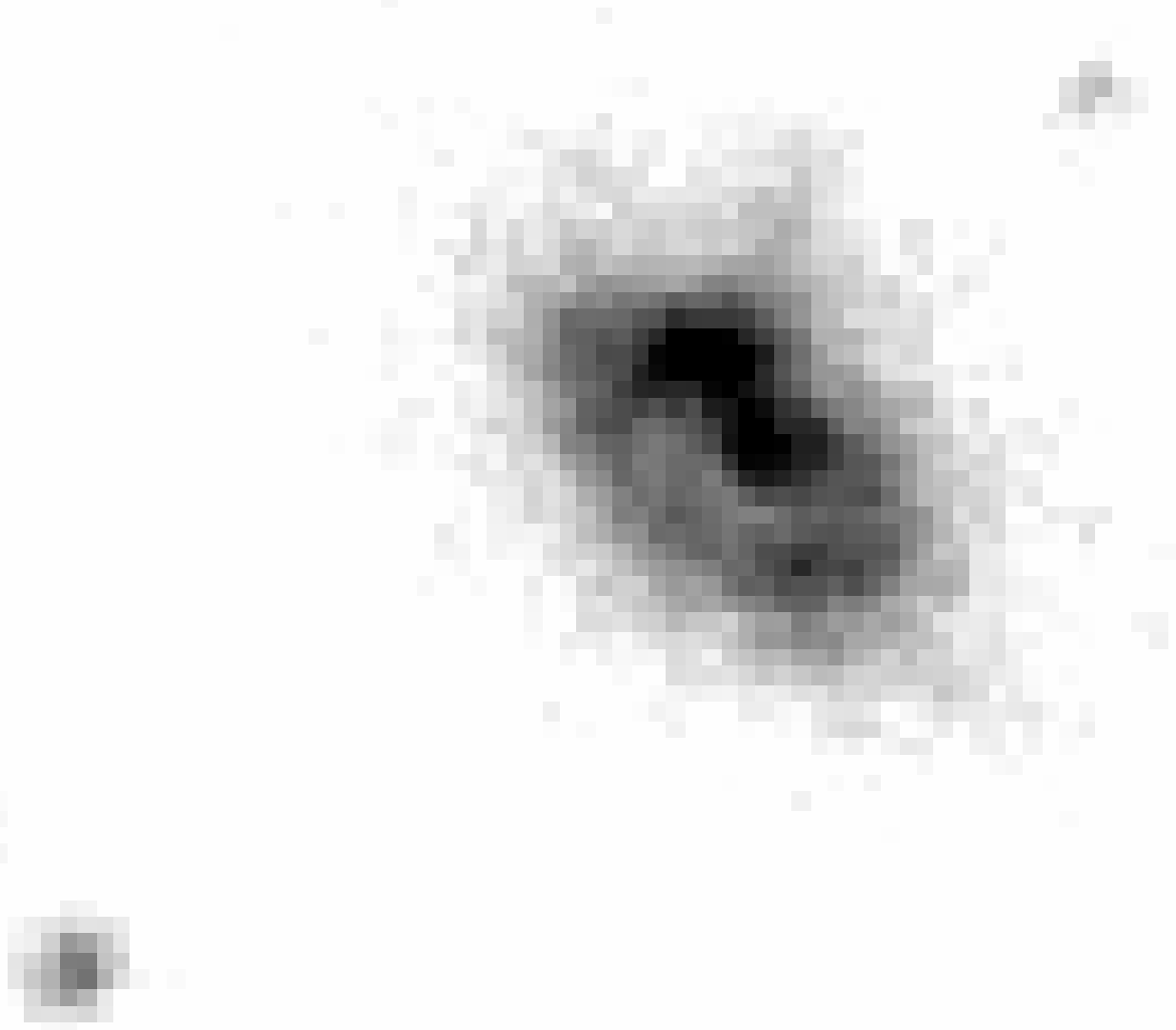,width=2truecm,clip=}%
\psfig{figure=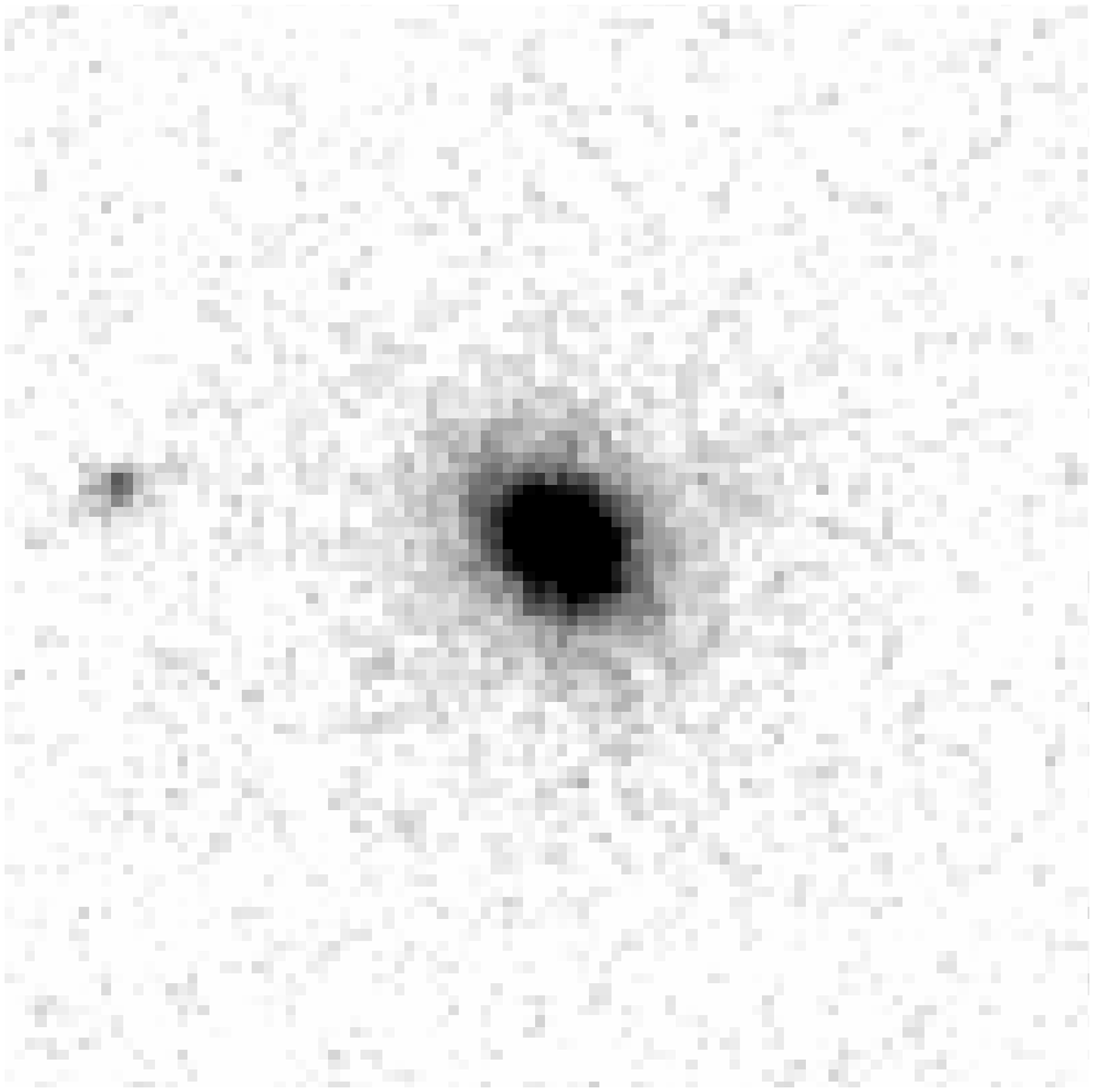,width=2truecm,clip=}%
\psfig{figure=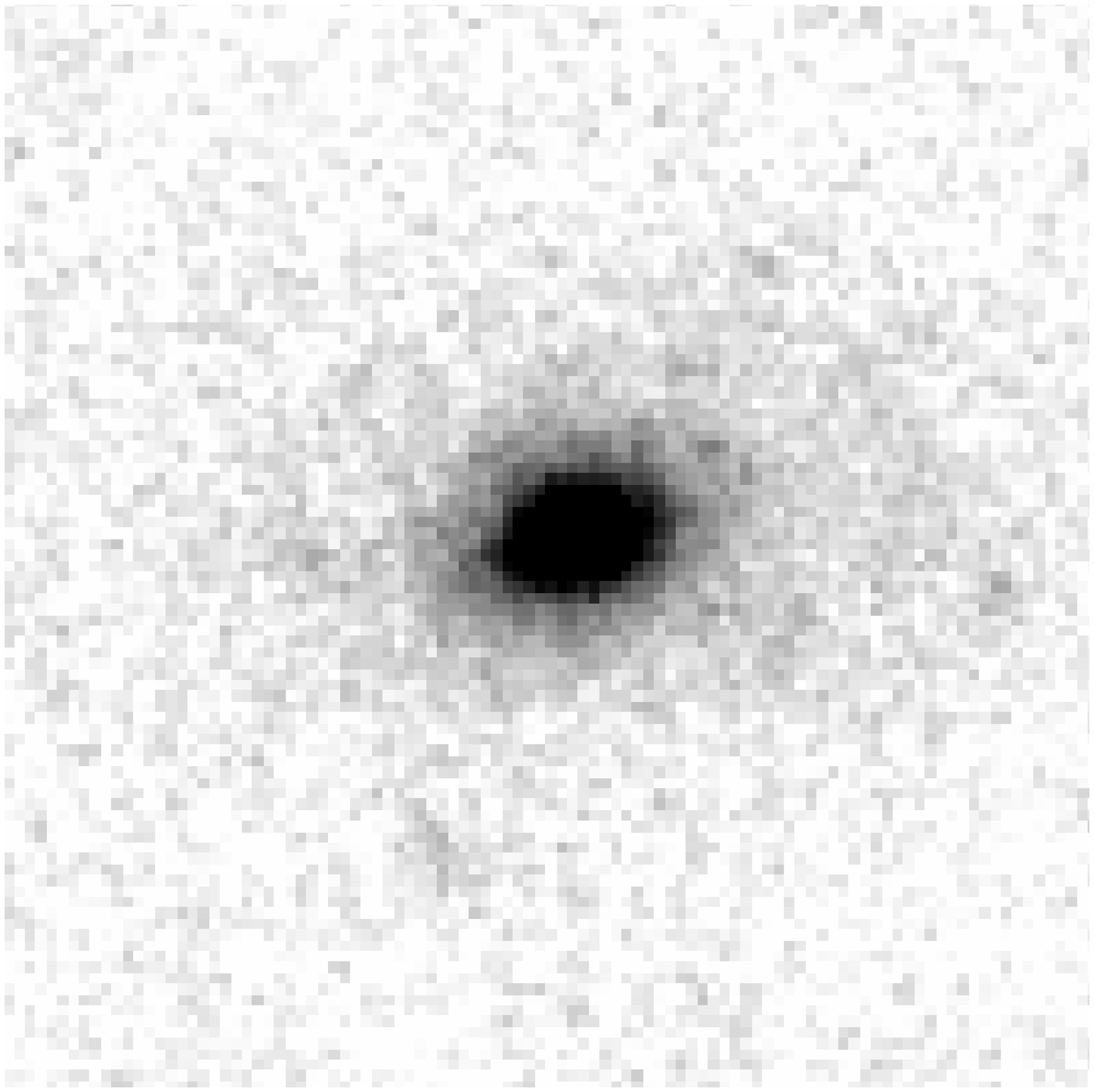,width=2truecm,clip=}%
\psfig{figure=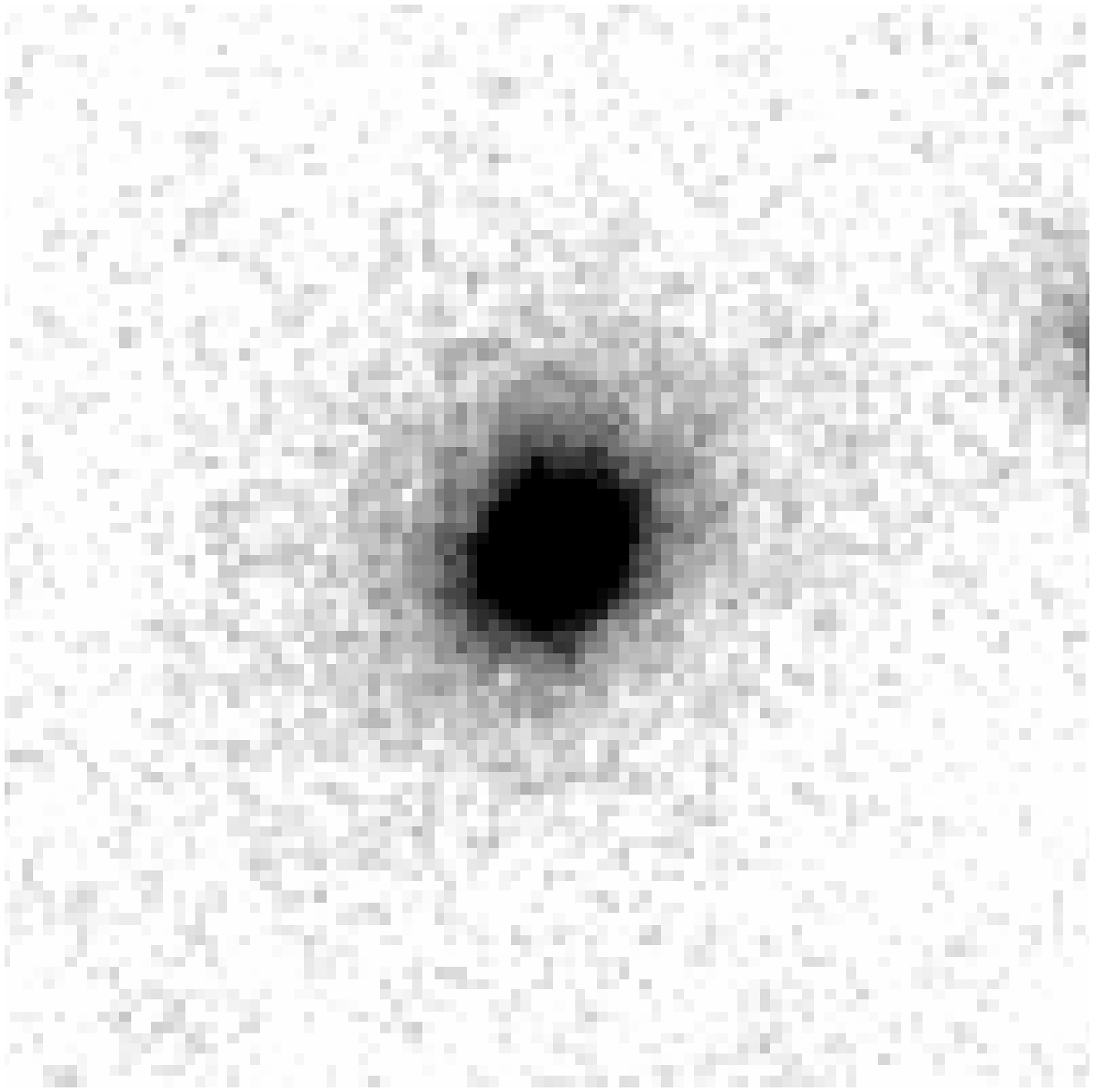,width=2truecm,clip=}%
}
\hbox{%
\psfig{figure=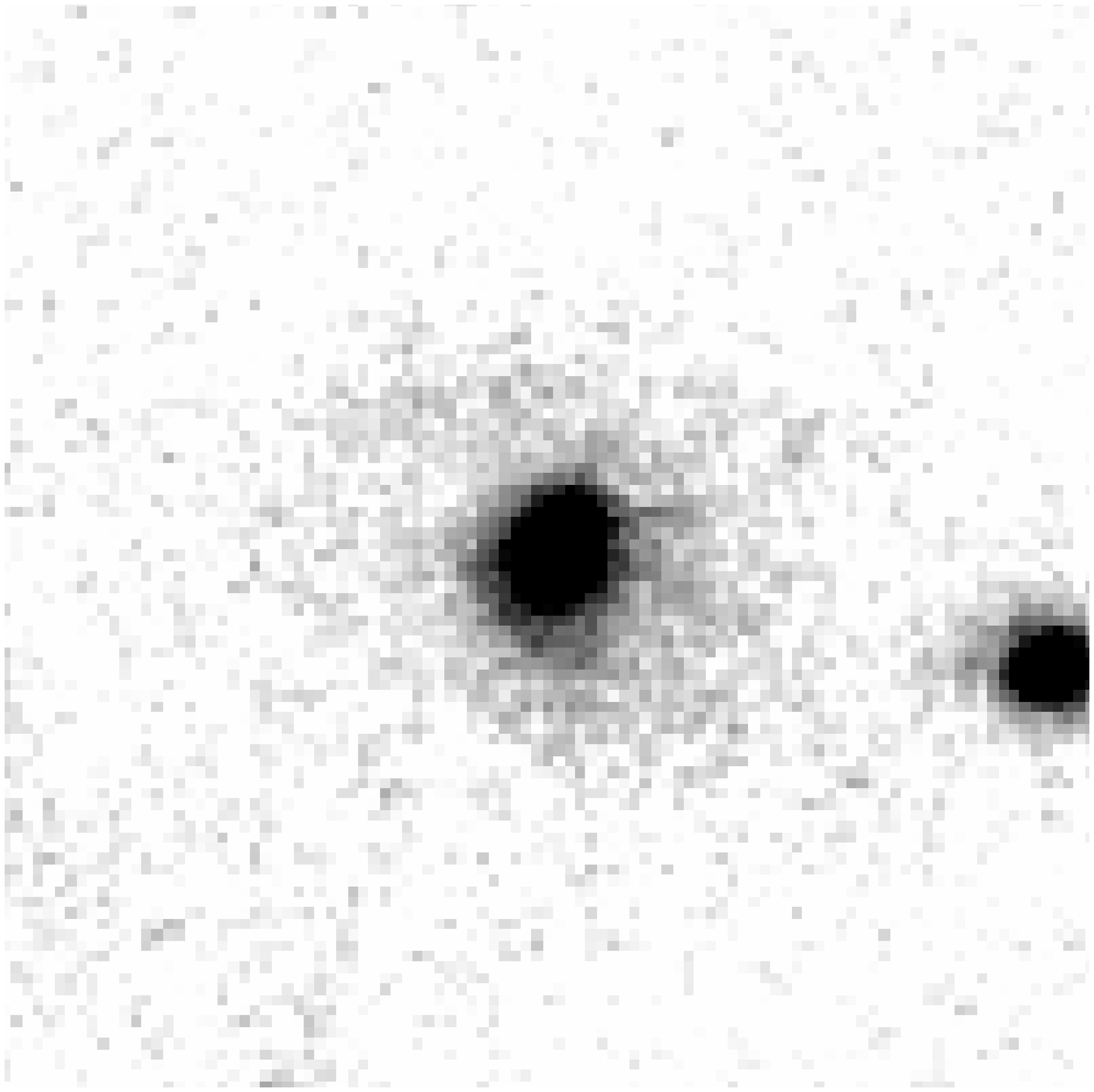,width=2truecm,clip=}%
\psfig{figure=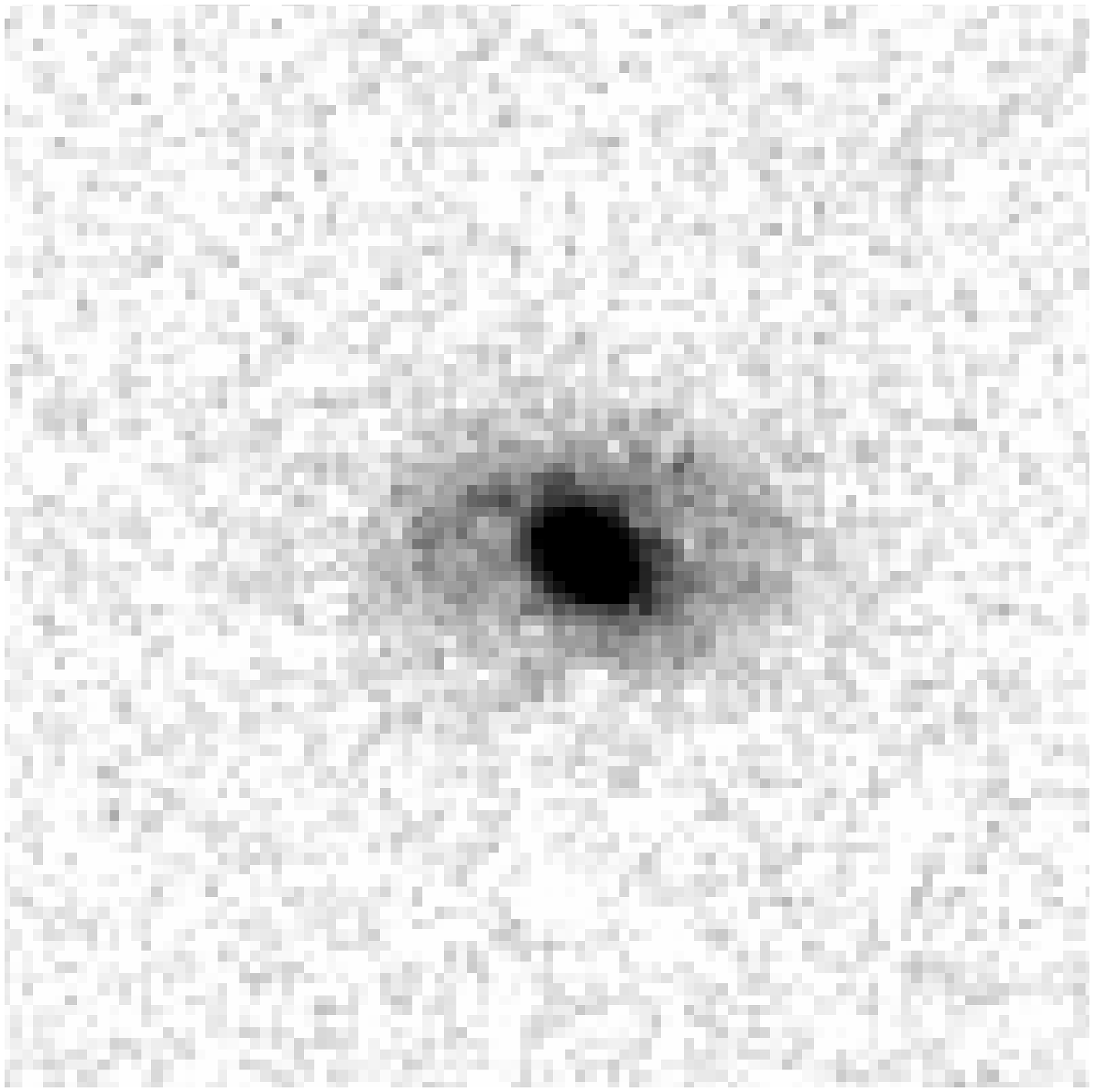,width=2truecm,clip=}%
}
\hbox{%
\psfig{figure=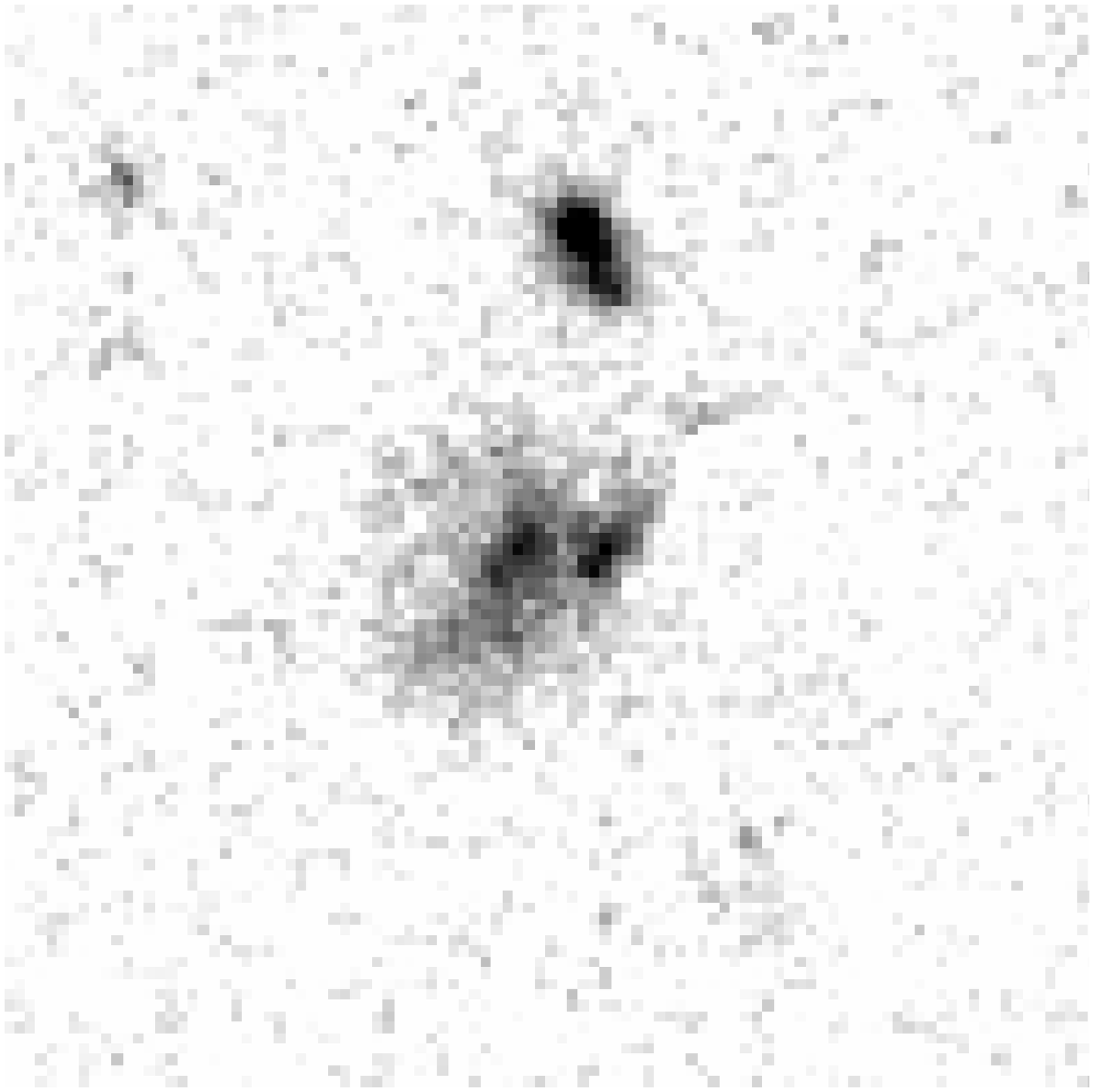,width=2truecm,clip=}%
\psfig{figure=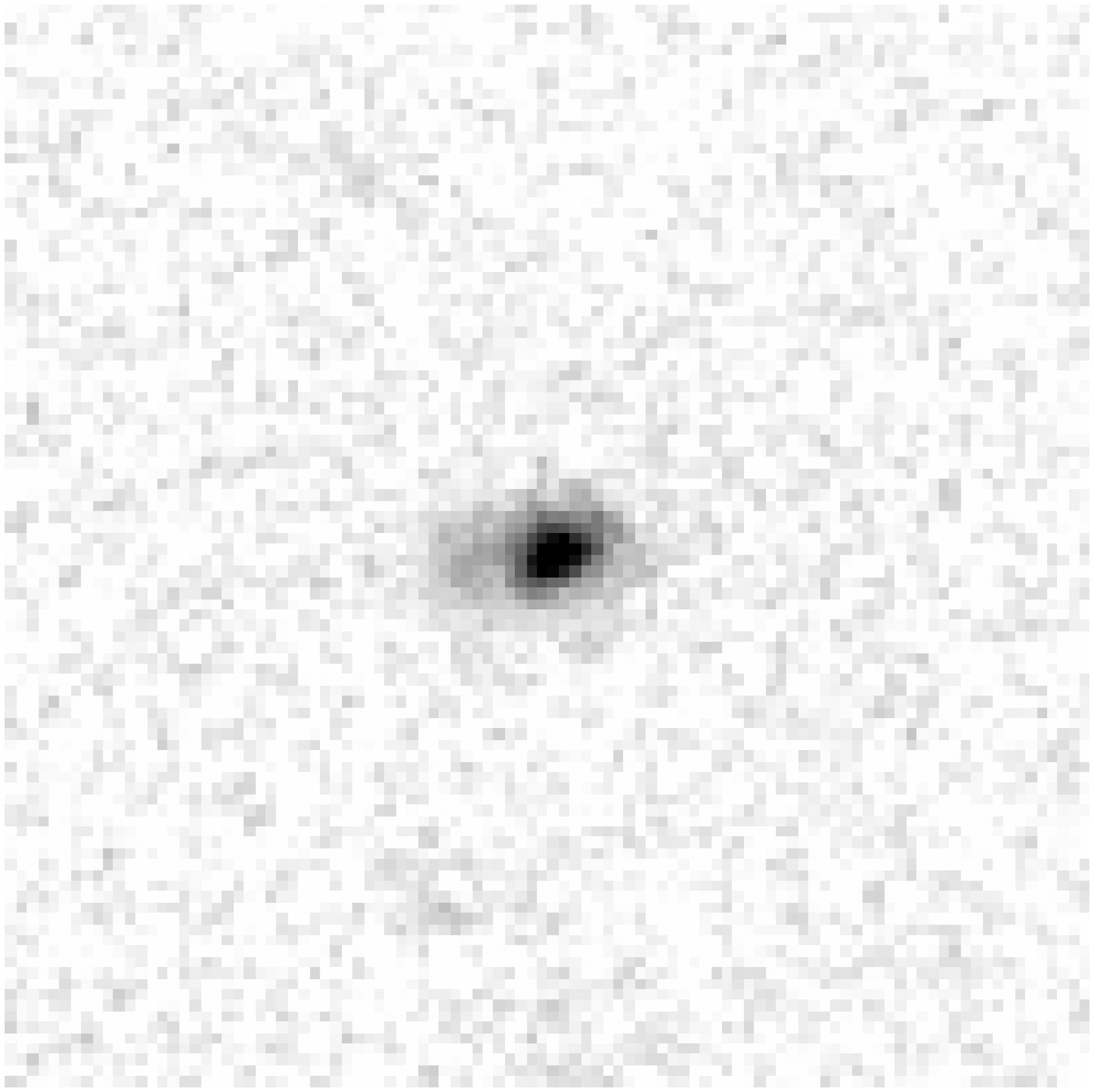,width=2truecm,clip=}%
\hskip 2 truecm
\psfig{figure=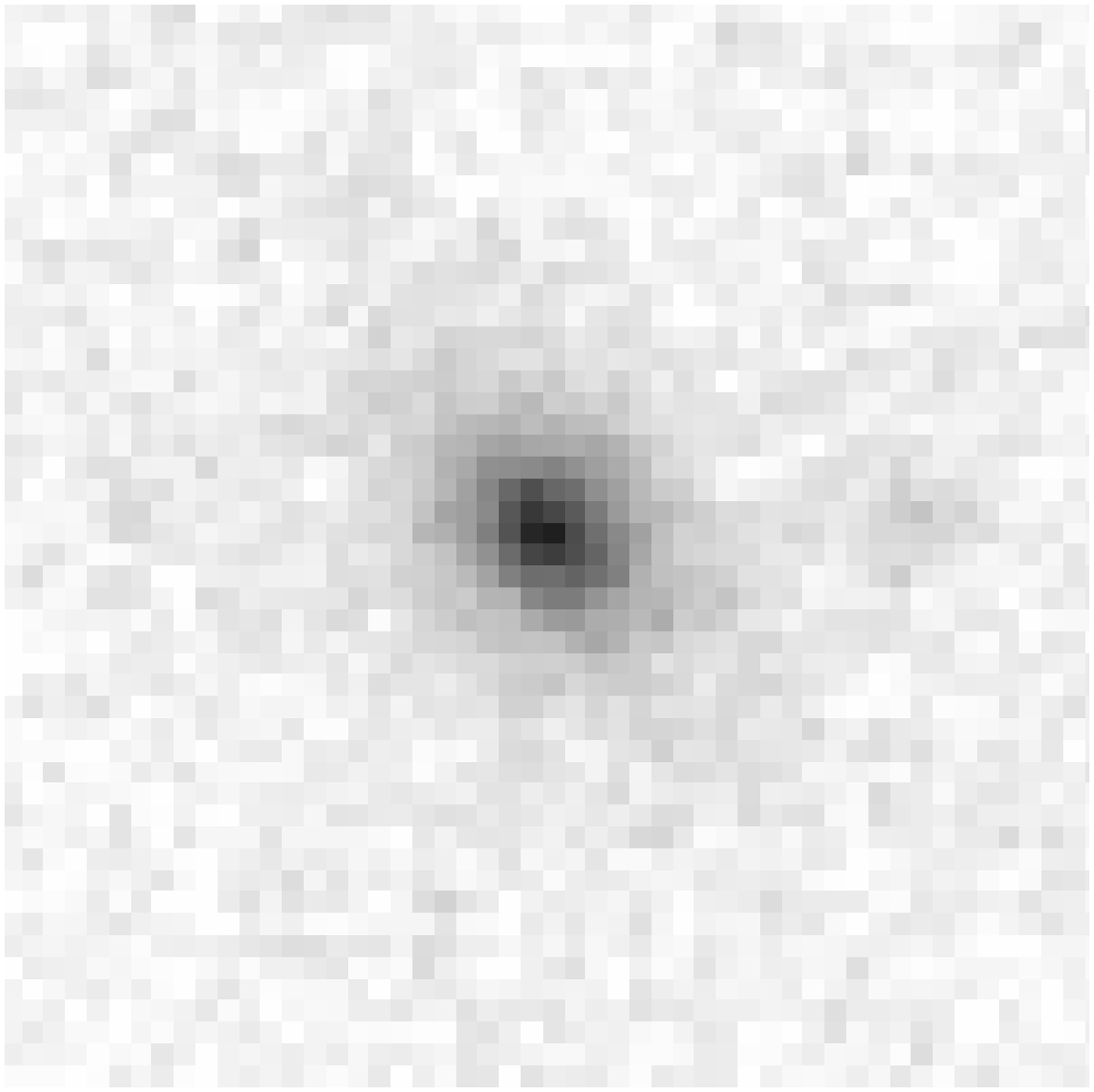,width=2truecm,clip=}%
}
\caption[h]{Gallery of HST images of spectroscopically
confirmed galaxies. The top row shows early-type galaxies.
The second to fourth rows show red ($I-z'>-0.12$ mag) late-type galaxies.
Last row shows blue ($I-z'<-0.12$ mag) late-type galaxies (two
left-most panels) or unclassified galaxy (righmost panel).
Each panel is $5''$ wide.}
\end{figure}

\section{The data}

The data set used here consists of a 2.4ks $I$ band image and a 1.4ks $z'$
band image obtained in July 2006, as part of the pre-imaging for deep
spectroscopy of RzCS 052, using the FOcal Reduction Spectrograph (FORS2)
imaging spectrograph at the European Southern
Observatory (ESO) 8.2m Very Large Telescope (VLT, progr. 075-A0175). 
FORS2 has a $7'\times7'$ field of view.
The typical seeing in these images varies between $0.5''$ and
$0.7''$, with 5$\sigma$ completeness limits (in $3''$ apertures, 
equivalent to a 24 kpc aperture), computed
as in Garilli, Maccagni \& Andreon 1999) of $I=24.5$
and $z'=24.0$ mag. The data were reduced and calibrated as usual except for the 
$z'$ band calibration, which is not part of
ESO's routine calibrations. This was bootstrapped from the CTIO $z'$
discovery image for this cluster presented in Andreon et al. (2007), in
turn calibrated using $z'$ standard stars (Smith et al. 2002). 
Both CTIO and FORS2 $z'$ have negligible colour terms (if any) with 
respect to the 
standard $z'$ system, as verified by observing a Sloan Digital Sky Survey (SDSS)
region with CTIO and with FORS2.

Additional high-resolution imaging for RzCS 052 was obtained from the HST
archive, which contains a 12ks $z$-band (F850LP) ACS image of this object (PID:
10496). The image was retrieved from the archive and processed through the {\tt
Multidrizzle} algorithm to remove hot pixels and cosmic rays (Koekemoer et al.
2002). The area sampled by ACS is about 2.2 Mpc$^2$, and 1/5th of the
FORS2 area. These data are used for our morphological study of galaxy 
populations in RzCS 052. Morphologies are estimated by eye (by SA),
as usual. For this reason, 20 \% of the estimated types are in error, on
average (Dressler et al. 1994, Couch et al. 1994, Andreon \& Davoust 1997).
Our magnitude limit for morphological typing, $z'=22.5$ mag, is conservatively set
at about 1.5 mag brighter than similar works (e.g., Blakeslee et al. 2006; 
Mei et al. 2006a,b) to limit the incidence of typing errors. 

Successful spectroscopy for around 54 galaxies, 21 of which turns out
to be cluster members (have a velocity offset less than
twice the computed cluster velocity dispersion), 
was obtained from VLT and Gemini GMOS spectra (see  
Andreon et al. 2007 for details). The morphology of the 16 confirmed members
that are in the ACS image is shown in Fig 1.

The spectroscopic sample is far from complete and, for our purposes, we need to
discriminate statistically against fore- and back-ground interlopers.  In order to
avoid systematic biases introduced by the use of heterogenous observational
data for cluster and control field (see e.g. Andreon et al. 2006a), we use two
$I$ and $z'$ FORS2 fields (progr. 073.A-0564) having similar depth and seeing to
our pre-imaging frames. In order to calibrate the $z'$-band photometry we make use
of a third archival program (072.C-0689) which acquired some $z'$-band imaging
of a region overlapping the SDSS (also including standard stars)
during the same nights.

Finally, in order to provide a local ($z \sim 0$) reference, we use our 
$u^*,g',r',z'$ observations
of Abell 496 at $z=0.032$ (Struble \& Rood 1999) taken (for another project)
in December 2005 with Megacam (Boulade et al. 2003) at CFHT under photometric
conditions.  Megacam has a field of $1^{\circ} \times 1^{\circ}$,  
and our images have seeing between $0.7''$ and $1.0''$ FWHM.
As for RzCS 052, the statistical discrimination of interlopers is
performed by using a control field, which observations have been interspersed
to Abell 496 observations. Data are reduced as usual and are at least 5 mag
deeper than needed for this project. 

We produce the object catalog and carry out photometry using the Source
Extractor software (SExtractor -- Bertin \& Arnout 1996). 
For RzCS052, photometry is carried out
in $3''$ (diameter) apertures, while colours are computed in $2''$ (diameter)
apertures. Abell 496 photometry uses, instead,
isophotal corrected magnitude (as ``total mag" proxy) and 3 arcsec aperture
for colours. Colours are corrected for minor differences (0.05 mag, or less) in
seeing, as in Andreon et al. (2004a).

Table 1 lists coordinates, $z'$ magnitude, $I-z'$ colour and
morphology of RzCS 052 spectroscopic members.

\begin{figure*}
\centerline{\psfig{figure=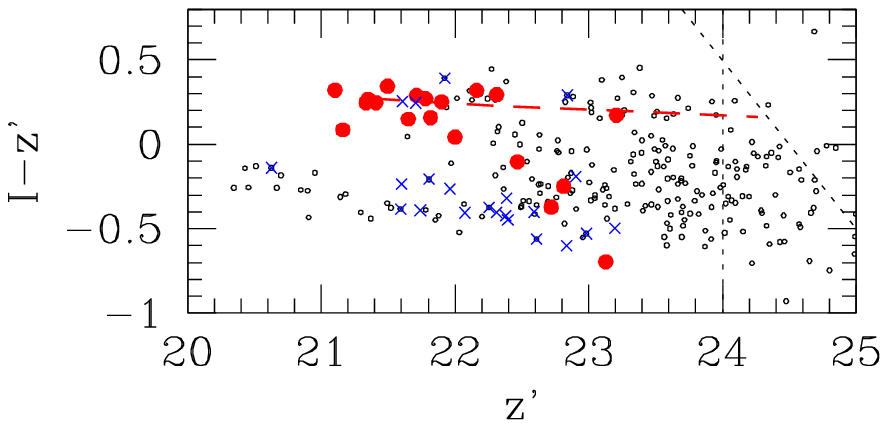,height=4.5truecm,clip=}
\psfig{figure=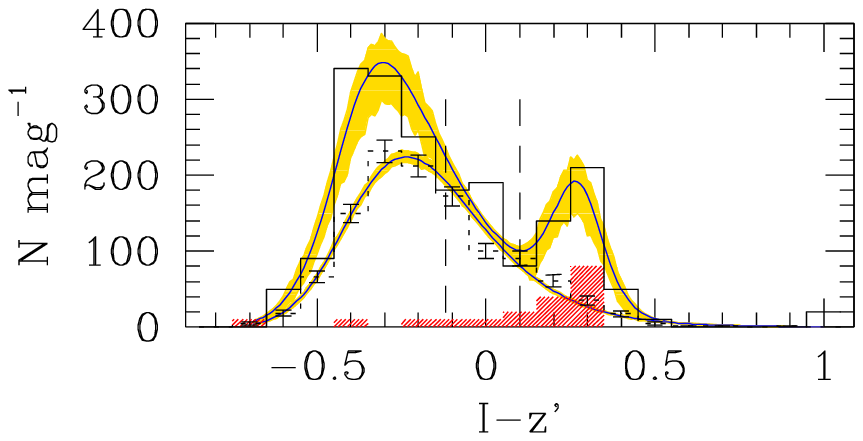,height=4.5truecm,clip=}}
\caption[h]{
{\it Left-hand panel:} Colour--magnitude diagram.
Galaxies with unknown membership within
$1.5'$ of the cluster center are shown as open circles,
known RzCS 052 members are marked with a (red) closed
circles. Spectroscopic non-members are marked with (blue) crosses.
The dashed (red) line is the expected colour--magnitude relation
at the cluster redshift,
from Kodama et al. (1997), without any parameter tuning.
{\it Right-hand panel:} Histogram of the colour distribution of galaxies
within $1.5'$ of the cluster center (solid histogram) and in the 72 arcmin$^2$
control fields (dashed -- normalized to the cluster area). The shaded histogram is
the colour distribution of confirmed cluster members. The shaded areas around the
smooth function show the 68 \% highest posterior credible intervals for cluster
and background colour distributions, modelling it as a 
mixture of a Pearson type
IV function (for background galaxies) and of two Gaussian's (for cluster galaxies),
following Bayesian methods (e.g. appendix B of Andreon et al. 2007). Finally, 
the two vertical (long-dashed) lines mark the boundary of the two 'red' definitions
considered in this work.
}
\end{figure*}

\input member_morpho.tex

\section{The red-sequence galaxies of RzCS 052}

The left-hand panel of Figure 2 shows the $I-z'$ vs. $z'$ color-magnitude diagram of
galaxies within $1.5'$ of the cluster center. This is equivalent to the
restframe $B-V$ vs. $V$. The familiar red sequence and the blue cloud of
star-forming galaxies can be clearly identified, even without marking the
spectroscopic members. It is easier to identify these features in the colour
histogram shown in the right-hand panel of Fig.~2. The narrow peak above the
normalized background at $I-z' \sim 0.35$ represents the red sequence, while
the broad feature at bluer colours is due to the blue cloud galaxies.

The line in the left panel of Figure 2 is the predicted color-magnitude relation from
Kodama \& Arimoto (1997), assuming a formation redshift of $z_f=5$ and total stellar
masses between 0.5 and 2 $\times 10^{11}\ M_{\odot}$. These models provide a
good fit not only to the present $I-z'$ relation, but also to the $R-z'$
relation in the discovery image and to the $R-z'$ color-magnitude relations
of the 18 $0.3 < z < 1.05$ clusters analyzed by Andreon et al (2004a). This
implies that red sequence galaxies in RzCS 052 appear to
be dominated by old stellar populations, similar to those in other high redshift
clusters, although the precise $z_f$ can vary between 2 and 11 depending
on model assumptions (see sec 3.3 of Andreon et al. 2004a). 

\begin{figure*}
\centerline{
\psfig{figure=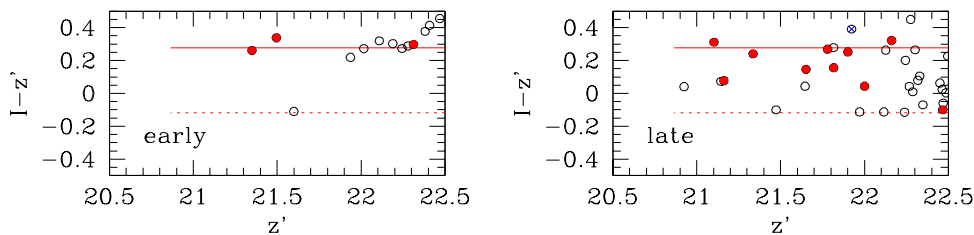,width=13truecm,clip=}%
}
\caption[h]{Colour-magnitude diagram of
galaxies in the HST field of view, having $I-z'>-0.12$ mag (marked by
a dotted line), 
for morphological early--types (left panel) and late-types
(right--panel) galaxies. 
Colours are corrected for the slope of the
colour--magnitude relation. From control field data,
we estimate that late--type galaxies with red colour
are a mix of interloper and member galaxies. 
Spectroscopic members and non-members are marked with (red) closed circles
and with (blue) crosses, respectively.
}
\end{figure*}

The intrinsic scatter about the color-magnitude relation provides a further
clue to the star formation history of these objects. We
calculate the dispersion for the 13
galaxies classified as early-type from the HST images and 
having $I-z'>-0.12$ mag (see Fig 3) using three methods: 
a) from the measured 
interquartile range without any outlier clipping; b) using the biweight
estimator of scale (Beers et al. 1990) and c) by fitting an Student-$t_4$ 
distribution, i.e. by Bayesian-fitting an overdispersed version of the Gaussian, in
order to account for outliers (Gelman et al. 2004). 
Once corrected for the photometric errors, as measured by
SExtractor, the intrinsic scatter is about $0.03-0.04$ mag,  
which is consistent with the local values 
(e.g. Bower, Lucey \&
Ellis 1992; Andreon 2003; Eisenhardt et al. 2007).
Following Mei et al. (2006a,b) we can transform this scatter into a mean
luminosity-weighted age corresponding to $z_f=2.5-3$ for both galaxy formation 
scenarios considered by Mei et al. (2006a,b). This is consistent with the usual
interpretation  that the majority of the stellar populations in the red sequence
galaxies were formed during rapid starbursts at high redshift.

It is important to emphasize that the the colour range over which the slope and
scatter of the CMR are computed must not be too narrow, otherwise this will
result in an artificial low value of the intrinsic scatter. For instance, 
for
a sample of galaxies uniformly distributed in colour, i.e. neither
old nor synchronized at all, are selected
in a 0.3 mag colour range (as in the ACS cluster survey, e.g. Mei et al. 
2006a,b), which has an effective (accounting for the colour-magnitude slope)
width of 0.2 mag, then 
the measured dispersion is $0.06$ mag ($=0.2/\sqrt{12}$).
Finding a
colour scatter that is small but compatible with an uniform distribution 
is, therefore, inconclusive about the star formation history in
samples severely colour pre-selected. 
Our estimate
of the colour scatter in RzCS 052 
is unaffected by the colour pre-selection, because 
we have considered galaxies up to $10\sigma$ away from the
colour--magnitude relation.

\begin{figure}
\psfig{figure=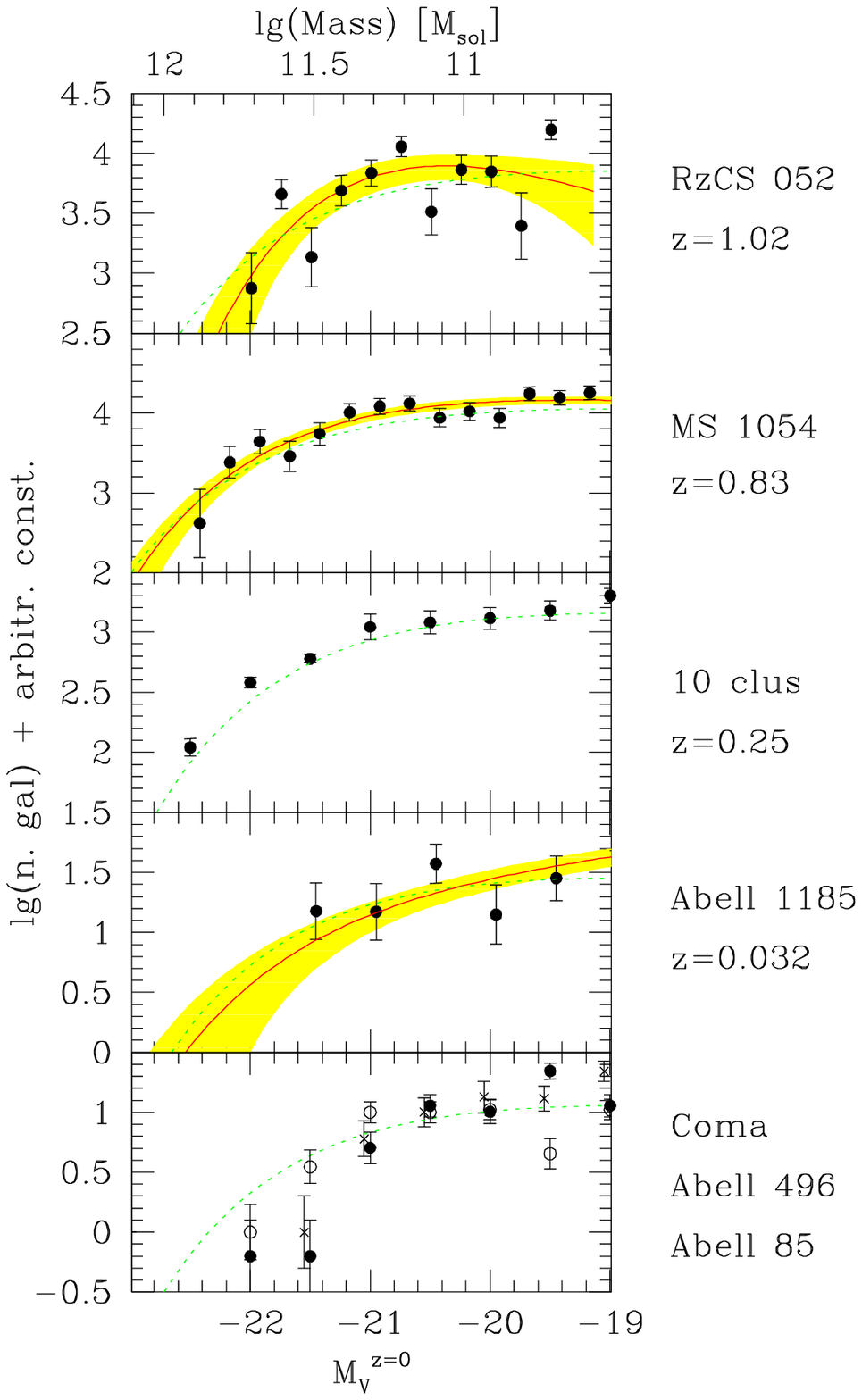,width=8truecm,clip=}
\caption[h]{Luminosity function of red galaxies in RzCS 052 (this paper),
MS1054 (Andreon 2006a), a composite luminosity function of 10 galaxies
at $z=0.25$ (Smail et al. 1988) and the local luminosity function of
Abell 1185 (Andreon et al. 2006a), Abell 1656 (crosses:
Secker et al. 1997; solid dots: Terlevich et al. 2001), and the
composite luminosity function of Abell 
496 and Abell 85 (open points), as derived by us from McIntosh et al. 
(2005) data. The points marks the LF as usually derived in
the astronomical literature, whereas the continuous (red) curve
and the shaded area mark LF and 68 \% confidence 
errors LFs derived following Andreon, Punzi \& Grado (2005) for
RzCS 052, and their Bayesian analogous for MS1054 and Abell 1185, as
described in those papers. 
The green line shows a reference luminosity function
with $\alpha=-1.0$ and $M^*_V=-21.3$ mag. All magnitudes
were transformed to $z=0$ (lower ascissa) or in mass (upper ascissa) 
using the procedure described in the text.
}
\end{figure}

Figure 3 shows the colour-magnitude diagram of galaxies classified
as early-type (left panel) and late-type (right panel).  The
difference in the colour scatter is notable between the two
morphological classes, even in the restricted color range studied.
Many of the red late-type galaxies are spectroscopic members
(marked by solid points), many, from control field observations,
are expected to be interlopers. The RzCS 052 red spirals may be
analogs of the anemic spirals encountered in low redshift clusters 
(e.g. Andreon 1996 and reference therein) or of
the massive disks of old stars at $z \sim 2$ described by Stockton, Canalizo
\& Maihara (2004). It is however unclear whether this represents evidence
of morphological evolution on the red sequence, because 
there are several
spirals also lying on the red sequence in the Coma cluster
(Andreon 1996,
Terlevich, Caldwell \& Bower 2001), even inside the cluster core.
We are presently investigating their nature (Covone et al., in preparation).

Even if the stellar populations of red sequence galaxies are uniformly old, it
is still possible that these objects are built via dissipationless mergers of
low mass spheroids (the so-called `red' or `dry' mergers). Pairs of close
spheroids have been identified in MS1054.3-03 by van Dokkum (1999) and Tran et al. (2005) or
in RDCS 0910+54 by Mei et al. (2006a). However, if red mergers are
important in the buildup of the red sequence, we should observe a gradual
increase in the mean mass of red galaxies with decreasing redshift. 
According to
De Lucia et al. (2006) galaxies more massive than $\sim 10^{11}$ $M_{\odot}$ 
accrete
about one half of their mass since $z=0.8$, which should make their
luminosity fainter than
the passively evolved local value by about $0.8$ mag at $z=1$ (and also
change the faint end slope, $\alpha$).

In order to test for the luminosity evolution,
we derive the rest-frame $V$ band luminosity and mass 
function of red galaxies in RzCS 
052. Red galaxies are defined as being redder than $I-z'>-0.12$, but we
note that adopting the valley between red and blue clouds ($I-z'=0.10$ mag,
see right panel of Figure 2) identical results are found.
Magnitudes were evolved to $z=0$
assuming a Bruzual \& Charlot (2003, BC03) single stellar population 
(SSP) 11.3 Gyr old at $z=0$ (i.e. $z_f=3$) and a Salpeter initial mass
function. We also compute a (stellar) mass function using the same BC03 model. 
The luminosity function  is calculated following Andreon, 
Punzi \& Grado (2005), accounting for the background contamination, estimated
from control fields, and performing the luminosity function computation
without binning the data and computing errors without the restrictive hypothesis 
that they (only) add up in quadrature.

Figure 4 shows the rest-frame $V$-band luminosity function of RzCS 052 and 
compares it with the luminosity functions of 16 clusters (see Figure for
details). There are 42 red statistical members in the RzCS 052 LF.

In our
sample of 16 clusters, we find no evidence of evolution in the luminosity (and
mass) function of red galaxies to $M^*+2$ once the  passive evolution of the
stellar populations is accounted for, as also found by many previous
studies (e.g. De Propris 1999, 2007, Andreon 2006a), but
for a sample dominated and not entirely composed by, red galaxies. We can
therefore {\it rule out} a significant contribution from mergers to the buildup
of the red sequence for galaxies with $M < M^*+2$ at $z \le 1$ in our
cluster sample as the majority of
their stellar mass is already in place. We emphasize once more that we assumed
a star formation history to convert luminosity to mass, but any assembly
history: the conversion from luminosity to stellar mass holds whether galaxies
formed monolithically or assembled hierarchically. A factor two increase in
mass due to mergers should move points 0.75 mag to the left in Fig. 4 going from
top to bottom.

Our finding, based on mass estimates derived from V-band rest frame 
magnitudes of a red-selected sample, is in good agreement with previous 
studies. For example, in the field, Wake et al. (2006)
find no evidence for
any additional evolution in the luminosity function of luminous red galaxies
beyond that expected from a simple passive evolution model.
Bundy, Ellis and Conselice (2005) find 
little evolution at the high mass end of the mass function of 
morphological early-type galaxies ($\approx 50$ \% of which with photometric 
redshifts): in particular the $M^*$ value of early-type galaxies is the 
same at the highest and lowest redshift bin.

One caveat is that our sample consists of just two high redshift
clusters (MS1054 and RzCS 052); however, in both cases, cluster galaxies are
more massive than the predictions of hierarachical assembly models and have
luminosities consistent with early assembly and no subsequent growth, via
merging or star formation. At very least, models
must accomodate both merger-driven and merger-free red sequence build up.

There is moderate evidence that RzCS 052 data points fluctuate too much around
the best fit model, i.e. that the Poisson model for the likelihood should be
updated with a more complex model that allows its variance to be different from
(larger than, in our case) its mean. However, such complex (and yet to be
determined) statistical model will not considerable move the RzCS 052
luminosity function to the right by about 0.75 mag, because the new model will
change the width of point error bars, not the value of points themselves, i.e.
it will not make our high redshift galaxies twice fainter and less
massive, which is what we need to reconcile data with a merger-driven
evolution. In particular, many red galaxies in the first magnitude bin are
spectroscopic members (see Fig 2), and any analysis intricacy 
can make these red massive galaxies disappear from the sample.

\section{The blue cloud galaxies in RzCS 052}

We now focus on the blue cloud galaxies. We characterize these objects via the
traditional blue fraction. Our definition of the blue fraction differs, however, from
the conventional definition by Butcher \& Oemler (1984) subsequently used
by most studies. We define a galaxy as blue if it is bluer than a BC03 model with
$z_f=3$ and $\tau=3.7$ Gyr. This galaxy will be bluer 
by 0.2 mag in $B-V$ than red sequence galaxies at the present
epoch (which would be a blue galaxy by the original definition). We also require
that this galaxy be brighter than exponential declining ($\tau$) or simple stellar 
populations (SSP) models having $M_V=-19.27$ mag at the present epoch (this avoids
objects momentarily brightened by starbursts).

\begin{figure*}
\centerline{\psfig{figure=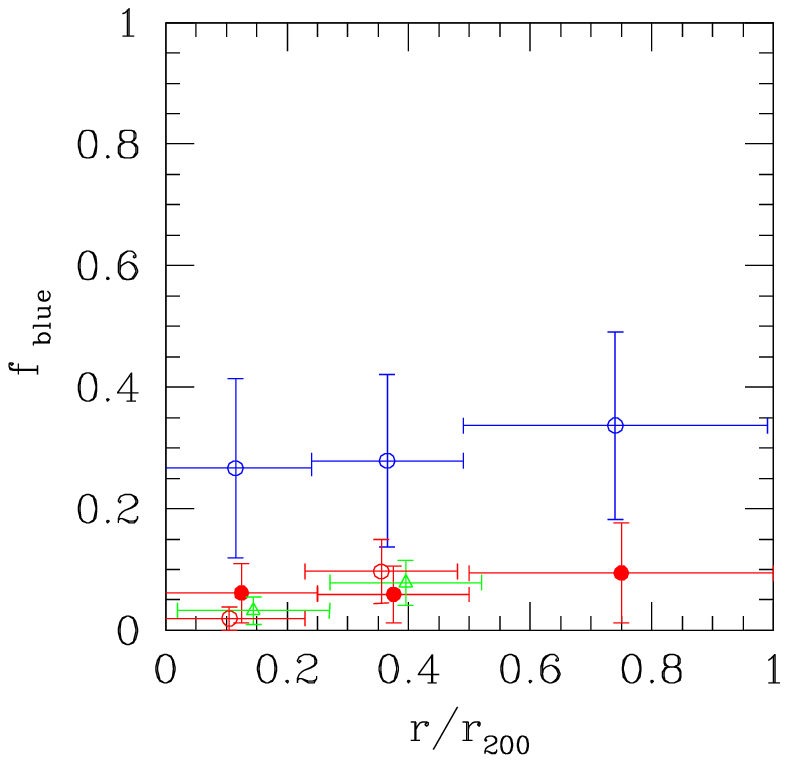,width=6truecm,clip=}%
\psfig{figure=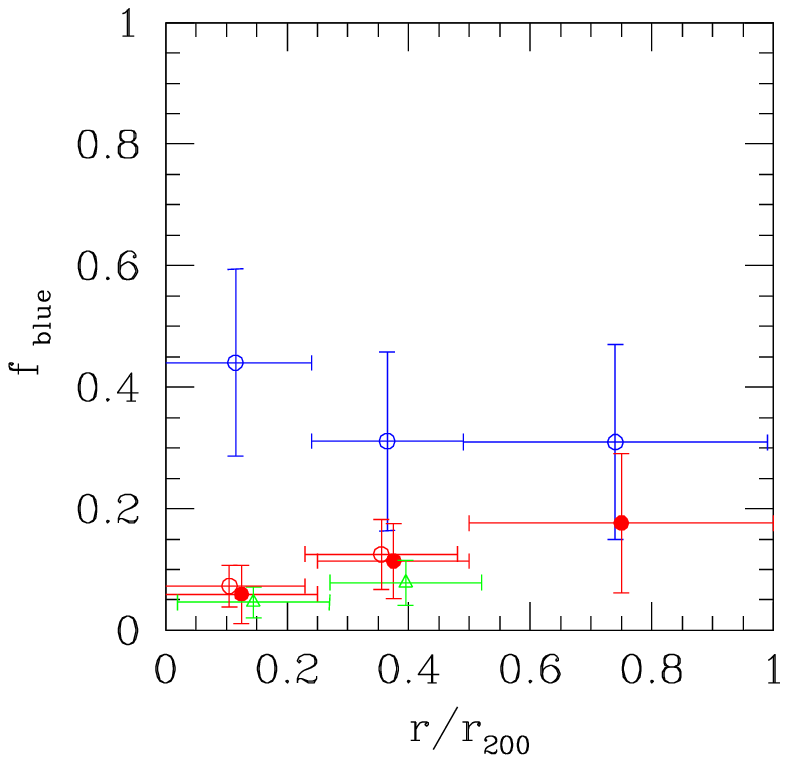,width=6truecm,clip=}%
}
\caption[h]{
Fraction of blue galaxies, as a function of clustercentric distance
expressed in units of virial radii: RzCS 052 (upper open blue points); Abell 496
from Megacam data (closed red points); Abell 496 from McIntosh et al. (2005)
photometry (lower open red points) and Abell 85 (green open triangles, also from
McIntosh et al. (2005) data). Fractions are computed for  mass-- (left panel) or
luminosity-- (right panel) selected samples. Points  are slightly displaced
horizontally for clarity. Error bars mark the posterior rms,  not the 68 \%
confidence interval.}
\end{figure*}

The rationale behind this choice is described in detail in Andreon et al. (2006b)
but in essence we attempt to account for the different star formation histories of
galaxies of different masses. To understand this better, let us consider a cluster
of galaxies formed at the same time, but with star formation histories (i.e., e-folding
times) dependent on galaxy mass, as in the original downsizing picture of Heavens
et al. (2004). It is obvious that the blue fraction in this cluster will change with
redshift even if there is no increase in the star formation rate due to the cluster
environment. Our approach attempts to identify `extra' star formation above and
beyond the increase in the blue fraction due to the younger mean age of the
Universe and the secular increase in the star formation rate with redshift.

We compute the blue
fraction accounting for background galaxies (i.e. in the line of sight,
and not belonging to the cluster) using our control fields 
following the Bayesian methods introduced in Andreon et al.
(2006b), for both a luminosity selected sample and a mass selected
sample. For the latter we adopt a mass threshold of $4 \ 10^{10}
M_{\odot}$, and a definition of mass given by the integral of the star
formation rate over  $0\le t \le \infty$. 

We computed the blue fraction for RzCS 052 members and also for two 
nearby galaxy clusters Abell 496 ($z=0.032$) and Abell 85 ($z=0.055$), 
in order to make a comparison with $z=0$. These two clusters have
velocity dispersions, and thus masses, almost identical to RzCS 052:
$\sigma_v=721^{+35}_{-20}$ (Rines et al. 2005) and 
$\sigma_v=692^{+55}_{-45}$ (Rines \& Diaferio 2006).
For Abell 496 we both used our Megacam data, and $U$ and $V$ catalogues 
published by McIntosh et al. (2005), also used to derive the blue
fraction of Abell 85. All virial radii 
are consistently derived from $\sigma_v$ using eq. 1 in 
Andreon et al. (2006b).

The two independent determinations of Abell 496 turn out to agree
each other and agree with the blue fraction of Abell 85 (see Fig. 5).

Figure 5 shows that the blue fractions of RzCS 052, Abell 496 and Abell 85
members does not show a large change with radius (both if the sample is luminosity 
or mass selected).
This result suggests that, whatever mechanism is changing the galaxy properties, 
it seems to affect in a similar way all the environments within
one virial radius. Therefore, mechanisms with a substantial different
efficiency at the center and at one virial radius are disfavored. Among
them, we mention tidal compression of galactic gas (e.g. Byrd \& Valtonen 1990),
that by interaction with the cluster potential can alter the star formation
rate, and tidal truncation of the outer galactic regions by the cluster
potential (e.g. Merritt 1983, 1984), that should remove the gas supply, after
eventually a last episode of star formation.

The fraction of blue galaxies is similar for mass-- and luminosity--  selected
samples in all three clusters, and is lower in nearby clusters than in RzCS 052. A
redshift-dependent blue fraction is favored, with respect to
redshift-independent one, with $\sim 10:1$ odds, i.e. there is positive
evidence of a change in the blue fraction between $z=0$ and $z=1$.
Abell 496 has  $f_b=0.03\pm0.03$ and $f_b=0.08\pm0.04$ for luminosity- and mass-
selected sample within $r_{200}$, whereas the respective values for   RzCS 052
are $f_b=0.28\pm0.10$ and $f_b=0.32\pm0.11$. 

The observed evolution in the blue
fraction from $z=1$ to $z=0$ does not take place because the fraction of red
galaxies decreases  with redshift (as shown above, the mass function of the red
galaxies is consistent with  pure passive evolution).  The change of the
blue fraction is therefore related to the blue population itself,
and it is not a reflex of a change in the red population. 
 
Due to the particular definitions of blue fraction adopted in this work, 
our finding is qualitatively different from
other (apparently similar) claims of a large fraction of blue galaxies at high
redshift (e.g. Butcher \& Oemler 1984) that, instead, does not account for 
the increase in the blue fraction due to the younger mean age of the
Universe and the secular increase in the star formation rate with redshift.
We are stating that we observe a blueing in excess to the one expected
in a younger universe, and thus
some more 
mechanism is working {\it in addition to the flow of the time} 
to make galaxies blue. 

Blue galaxies might merge to form galaxies on the red
sequence, but this would conflict with both the observed evolution of the mass
function of galaxies on the red sequence presented in the previous section
and the constant mass function of galaxies in
32 clusters, including RzCS 052, at $0.29<z<1.25$, which is the unique
scenario not excluded by Andreon (2006a), and the 
preferred scenario in other studies (e.g. De Propris et al. 1999; 2007).
The only remaining possibility is that the blue galaxy evolution is
faster than our library of models used to describe their star
formation history ($\tau$ models), i.e. that blue galaxies in RzCS 052
are experimenting a vigorous 
star formation history, possibly involving starbursts (perhaps short-lived).
This same
suggestion is pointed out by the analysis of the rest-frame blue LF of 24
clusters at $0.3<z<1.05$ in Andreon et al. (2004a): $M^*$ values are not 
aligned on
a common evolutionary track (see their Fig 8), but scatter off as much as 1 mag,
a clear indication of a complex star formation history.

Since RzCS 052 has a low x-ray emission (see Andreon et al.
2007)
than other clusters with an x-ray bright intracluster medium,
gas-related quenching mechanisms, such as ram pressure stripping (e.g. Gunn
\& Gott 1972; Byrd \& Valtonen 1990; Quilis, Moore \& Bower 2000),
turbulent and viscous stripping (Nulsen 1982), thermal evaporation (Cowie \&
Songaila 1977) and pressure-triggered star formation (Dressler \& Gunn
1983), will be much less efficient (e.g. Treu et al. 2003), making unlikely that something that
lacks, the gas, play a role in making RzCS 052 galaxies bluer than
expected. On the other side, an opposite conclusion can be drawn: the
relative lack of gas in RzCS 052 (and therefore less effective quenching)
compared to the x-ray bright clusters allows for more extended and
vigorous star formation and therefore RzCS 052 has a large blue fraction
because the lack of the gas that ultimately ends star formation in
infalling galaxies.

\section{Conclusions}

We presented new results on morphology, mass
assembly history, role of environment and colour bimodality of
galaxies in the colour selected, modest x-ray emitter, cluster RzCS 052 at
$z=1.02$, as derived from VLT, HST, and  optical 
data, supplemented by a coarser analysis of a large sample of about
45 clusters, from $z=0$ to $z=1.22$ (16 in the context of
the evolution of red galaxies shown in Fig. 4, and 32 used to
constraint the fate of the blue population in Sec 4).

We found that the colour distribution of RzCS 052 is bimodal.
Analysis of the morphological mix, slope and intercept of
the colour-magnitude relation, and of the mass function shows 
that RzCS 052 red galaxies differs only by age from their local 
counterparts and that mergers play a minor role in building 
massive (down to~$M^*+2$) red galaxies in studied clusters, 
from $z=1$ to today. 

The situation is remarkably different for blue galaxies. The blue
fraction, once accounted for the younger age of stellar populations
at high redshift and for the higher star formation rate there, is
larger in RzCS 052 than in nearby similar clusters, highlighting
perhaps for the first time that something, in addition to
the flow of the time, is making galaxies bluer at high redshift. 

Mergers are unlikely to be the driver of the observed colour
evolution between $z=1$ and $z=0$, because of the measured constancy of
the mass function, as derived from Spitzer photometry of 32
clusters in Andreon (2006). 
Mechanisms requiring a substantial intracluster medium,
such as ram pressure stripping, are ruled out as well as direct
driver, because of the very modest x-ray emission in RzCS 052. Mechanisms
with a substantial different efficiency at the center and
at one virial radius are strongly disfavored, because of the
observed constant blue fraction.

\section*{Acknowledgments}

SA would like to thank Oscar Straniero, Diego de Falco, Nelson Caldwell,
Giovanni Covone
and Kodama Tadayuki for their help during the preparation of this paper.
We thank the referee for his/her careful advices.
It is also a pleasure to acknowledge the hospitality of the University
of Bristol, where early discussions about this work took place. 
SA acknowledge financial contribution from contract ASI-INAF I/023/05/0.

We acknowledge observations taken at 
ESO (75.A-0175), 
HST (progr. 10496), 
CFHT (2005BD96).

\bsp

\label{lastpage}

\end{document}

%% file: member_morpho.tex
\begin{table}
\caption{J2000 coordinates, $z'$ magnitude, $I-z'$ colour and
morphology of RzCS 052 spectroscopic members}
\begin{tabular}{ccccl}
\hline
 RA  &  DEC  & $z'$ & $I-z'$ & morph. \\
\hline
\hline                                                   
02:21:36.21 & -03:24:56.0 & 21.71 & 0.29  &  \\       
02:21:37.08 & -03:24:28.4 & 21.41 & 0.25  &  \\    
02:21:37.60 & -03:21:38.0 & 22.00 & 0.04  & late \\			    
02:21:38.85 & -03:23:40.7 & 22.47 & -0.10 & late \\			  
02:21:39.60 & -03:22:00.9 & 21.50 & 0.34  & early \\			
02:21:40.32 & -03:19:03.4 &	  &	  &  \\ 	     
02:21:40.46 & -03:18:35.6 &	  &	  &  \\   
02:21:41.13 & -03:24:41.2 & 23.13 & -0.70 &  \\ 	   
02:21:41.73 & -03:23:35.2 & 22.72 & -0.37 & compact \\  	      
02:21:42.04 & -03:21:54.1 & 21.90 & 0.25  & late \\    
02:21:42.14 & -03:20:07.0 & 21.65 & 0.15  & late \\    
02:21:42.52 & -03:22:43.6 & 22.81 & -0.25 & late \\    
02:21:42.81 & -03:22:48.8 & 22.16 & 0.32  & late \\    
02:21:43.15 & -03:21:15.2 & 21.16 & 0.09  & late \\    
02:21:43.87 & -03:21:06.0 & 21.35 & 0.27  & early \\   
02:21:43.96 & -03:20:27.9 & 23.21 & 0.17  & late \\    
02:21:44.85 & -03:22:04.3 & 21.10 & 0.32  & late \\    
02:21:44.90 & -03:21:44.5 & 21.34 & 0.25  & late \\    
02:21:45.21 & -03:21:25.5 & 21.78 & 0.27  & late \\    
02:21:45.24 & -03:20:44.3 & 22.31 & 0.29  & early \\   
02:21:48.33 & -03:20:48.6 & 21.82 & 0.16  & late \\    
\hline 
\end{tabular}  
\hfill \break 
Some galaxies have no morphological type or colour because they are outside
the ACS or FORS2 field of view. 
\end{table}